# Comparison of Analytical Solutions of DGLAP Evolution Equations

R. Rajkhowa

Physics Department, T. H. B. College, Jamugurihat, Sonitpur, Assam.

E-mail:rasna.rajkhowa@gmail.com

**Abstract**

We explain particular, unique, approximate solutions of the Dokshitzer-Gribov-Lipatov-Altarelli-Parisi (DGLAP) evolution equations and also solutions of DGLAP evolution equations by using regge behaviour of structure functions and method of characteristic for $t$ and $x$-evolutions of singlet and non-singlet structure functions in leading order (LO) and next-to-leading order (NLO). Hence $t$-evolution of deuteron, proton, neutron and difference of proton and neutron and $x$-evolution of deuteron, proton and neutron structure functions in LO and NLO at low-$x$ from DGLAP evolution equations. The results of $t$ and $x$-evolutions are compared with experimental data and global parameterization in different kinematics region. We also compare the solutions of DGLAP evolution equations among themselves.

PACS No.: 12.38.Bx, 12.39-x, 13.60Hb.

Keywords: Particular solution, Complete solution, Unique solution, Approximate Solution, Regge behaviour, Characteristic method, Altarelli-Parisi equation, Structure function, Low-$x$ physics.

## 1. Introduction

Among different evolution equations, up till now Dokshitzer-Gribov-Lipatov-Altarelli-Parisi (DGLAP) [1-4] evolution equations are most successful and major tools to study the structure functions of hadrons and ultimately structure of matter, ultrahigh-energy cosmic rays. Various methods, like particular, unique, approximate, characteristic, regge and brute-force, Laguerre-polynomial, Mellin-transformation etc. methods have been developed for the analytical and numerical solution. In this paper, we are concentrate our work mainly in different analytical solutions of DGLAP evolution equations in leading order (LO), next-to-leading order (NLO) and compare them particularly by focusing on the numerical accuracy, approximation and better fitness of results with experimental data and global parameterization in different kinematics region. Here, we explain particular, unique, approximate solutions of the DGLAP evolution equations and also solutions of DGLAP evolution equations by using regge behaviour of structure functions and method of characteristic for $t$ and $x$-evolutions of singlet and non-singlet structure functions in leading order (LO) and next-to-leading order (NLO). Hence *t*-evolution of deuteron, proton, neutron, difference of proton and neutron and *x*-evolution of deuteron and proton structure functions in LO and NLO at low-*x* from DGLAP evolution equations. The results of *t* and *x*-evolutions are compared with experimental data in different kinematics region. We also compare the solutions of DGLAP evolution equations among themselves.

## 2. Theory

Though the necessary theory has been discussed elsewhere [5-13], here we mention some essential steps for clarity. The DGLAP evolution equations with splitting functions [14-16] for singlet and non-singlet structure functions in LO and NLO are in the standard forms [5-13, 17]

$$\frac{\partial F_2^S(x,t)}{\partial t} - \frac{\alpha'_s(t)}{2\pi}\left[\frac{2}{3}\{3+4\ln(1-x)\}F_2^S(x,t) + \frac{4}{3}\int_x^1\frac{dw}{1-w}\{(1+w^2)F_2^S\left(\frac{x}{w},t\right) - 2F_2^S(x,t)\}\right]$$

$$+ n_f \int_x^1 \{w^2 + (1-w)^2\}G(\frac{x}{w},t)dw]] = 0, \qquad (1)$$

$$\frac{\partial F_2^{NS}(x,t)}{\partial t} - \frac{\alpha'_s(t)}{2\pi}\left[\frac{2}{3}\{3+4\ln(1-x)\}F_2^{NS}(x,t) + \frac{4}{3}\int_x^1\frac{dw}{1-w}\{(1+w^2)F_2^{NS}\left(\frac{x}{w},t\right) - 2F_2^{NS}(x,t)\}\right] = 0, \qquad (2)$$

for LO. And

$$\frac{\partial F_2^S(x,t)}{\partial t} - \frac{\alpha_s(t)}{2\pi}\left[\frac{2}{3}\{3+4\ln(1-x)\}F_2^S(x,t) + \frac{4}{3}\int_x^1\frac{dw}{1-w}\{(1+w^2)F_2^S\left(\frac{x}{w},t\right) - 2F_2^S(x,t)\}\right.$$

$$+ n_f \int_x^1 \{w^2 + (1-w)^2\}G(\frac{x}{w},t)dw]] - \left(\frac{\alpha_s(t)}{2\pi}\right)^2\left[(x-1)F_2^S(x,t)\int_0^1 f(w)dw + \int_x^1 f(w)F_2^S(\frac{x}{w},t)dw\right.$$

$$+ \int_x^1 F_{qq}^S(w)F_2^S(\frac{x}{w},t)dw + \int_x^1 F_{qg}^S(w)G(\frac{x}{w},t)dw] = 0, \qquad (3)$$

$$\frac{\partial F_2^{NS}(x,t)}{\partial t} - \frac{\alpha_s(t)}{2\pi}\left[\frac{2}{3}\{3+4\ln(1-x)\}F_2^{NS}(x,t) + \frac{4}{3}\int_x^1\frac{dw}{1-w}\{(1+w^2)F_2^{NS}\left(\frac{x}{w},t\right) - 2F_2^{NS}(x,t)\}\right]$$

$$- \left(\frac{\alpha_s(t)}{2\pi}\right)^2\left[(x-1)F_2^{NS}(x,t)\int_0^1 f(w)dw + \int_x^1 f(w)F_2^{NS}\left(\frac{x}{w},t\right)dw\right] = 0 \qquad (4)$$

for NLO, where

$$\alpha'_s(t) = \frac{4\pi}{\beta_0 t}, \quad \alpha_s(t) = \frac{4\pi}{\beta_0 t}\left[1 - \frac{\beta_1 \ln t}{\beta_0^2 t}\right], \quad \beta_0 = \frac{33-2N_f}{3} \text{ and } \beta_1 = \frac{306-38N_f}{3},$$

$N_f$ being the number of flavours. Here,

$$f(w) = C_F^2 [P_F(w) - P_A(w)] + \frac{1}{2} C_F C_A [P_G(w) + P_A(w)] + C_F T_R N_f P_{N_F}(w),$$

$$F_{qq}^S(w) = 2 C_F T_R N_f F_{qq}(w)$$

and $\quad F_{qg}^S(w) = C_F T_R N_f F_{qg}^1(w) + C_G T_R N_f F_{qg}^2(w).$

The explicit forms of higher order kernels are [14-16]

$$P_F(w) = -2\left(\frac{1+w^2}{1-w}\right) \ln w \ln(1-w) - \left(\frac{3}{1-w} + 2w\right) \ln w - \frac{1}{2}(1+w)\ln^2 w - 5(1-w),$$

$$P_G(w) = \frac{1+w^2}{1-w}\left(\ln^2 w + \frac{11}{3} \ln w + \frac{67}{9} - \frac{\pi^2}{3}\right) + 2(1+w)\ln w + \frac{40}{3}(1-w),$$

$$P_{N_F}(w) = \frac{2}{3}\left[\frac{1+w^2}{1-w}\left(-\ln w - \frac{5}{3}\right) - 2(1-w)\right],$$

$$P_A(w) = 2\left(\frac{1+w^2}{1-w}\right) \int_{w/(1+w)}^{1/(1+w)} \frac{dk}{k} \ln\frac{1-k}{k} + 2(1+w)\ln w + 4(1-w),$$

$$F_{qq}(w) = \frac{20}{9w} - 2 + 6w - \frac{56}{9}w^2 + \left(1 + 5w + \frac{8}{3}w^2\right)\ln w - (1+w)\ln^2 w,$$

$$F_{qg}^1(w) = 4 - 9w - (1-4w)\ln w - (1-2w)\ln^2 w + 4\ln(1-w)$$
$$+ \left[2\ln^2\left(\frac{1-w}{w}\right) - 4\ln\left(\frac{1-w}{w}\right) - \frac{2}{3}\pi^2 + 10\right] P_{qg}(w)$$

and

$$F_{qg}^2(w) = \frac{182}{9} + \frac{14}{9}w + \frac{40}{9w} + \left(\frac{136}{3}w - \frac{38}{3}\right)\ln w - 4\ln(1-w) - (2+8w)\ln^2 w$$
$$+ \left[-\ln^2 w + \frac{44}{3}\ln w - 2\ln^2(1-w) + 4\ln(1-w) + \frac{\pi^2}{3} - \frac{218}{3}\right] P_{qg}(w) + 2 P_{qg}(-w) \int_{w/(1+w)}^{1/(1+w)} \frac{dz}{z} \ln\frac{1-z}{z},$$

where, $P_{qg}(w) = w^2 + (1-w)^2$, $C_A$, $C_G$, $C_F$, and $T_R$ are constants associated with the color SU(3) group and $C_A = C_G = N_C = 3$, $C_F = (N_C^2 - 1)/2N_C$ and $T_R = 1/2$. $N_C$ is the number of colours.

Using the variable $u = 1-w$ and Taylor expansion method [19], singlet structure function $F_2^S(x/w, t)$ and gluon structure function $G(x/w, t)$ can be approximated for small-$x$ as

$$F_2^S(x/w,t) \cong F_2^S(x,t) + x \cdot \frac{u}{1-u} \frac{\partial F_2^S(x,t)}{\partial x}. \tag{5}$$

Similarly, $G(x/w, t)$ can be approximated for small-$x$ as

$$G(x/w,t) \cong G(x,t) + x \cdot \frac{u}{1-u} \frac{\partial G(x,t)}{\partial x}. \tag{6}$$

Using equations (5) and (6) in equation (1) and performing *u*-integrations we get

$$\frac{\partial F_2^S(x,t)}{\partial t} - \frac{\alpha_s'(t)}{2\pi}\left[A_1(x)F_2^S(x,t) + A_2(x)G(x,t) + A_3(x)\frac{\partial F_2^S(x,t)}{\partial x} + A_4(x)\frac{\partial G(x,t)}{\partial x}\right] = 0. \quad (7)$$

Where, $A_1(x)$, $A_2(x)$, $A_3(x)$ and $A_4(x)$ are some functions of *x* [5].

We assume [5-13]

$$G(x, t) = K(x)\, F_2^S(x, t). \quad (8)$$

where $K(x)$ is a parameter to be determined from phenomenological analysis and we assume $K(x) = K$, $ax^b$ or $ce^{dx}$ where $K$, $a$, $b$, $c$ and $d$ are constants. Though we have assumed some simple standard functional forms of $K(x)$, yet we can not rule out the other possibilities.

Therefore equations (7) becomes

$$\frac{\partial F_2^S(x,t)}{\partial t} - \frac{A_f}{t}\left[L_1(x)F_2^S(x,t) + L_2(x)\frac{\partial F_2^S(x,t)}{\partial x}\right] = 0. \quad (9)$$

Here,

$$L_1(x) = A_1(x) + K(x)A_2(x) + A_4(x)\frac{\partial K(x)}{\partial x}, \quad L_2(x) = A_3(x) + K(x)A_4(x) \text{ and } A_f = 4/(33-2n_f).$$

The general solutions [19-20] of equation (9) are

$$U\left(x, t, F_2^S\right) = t\exp\left[\frac{1}{A_f}\int\frac{1}{L_2(x)}dx\right] \text{ and } V\left(x, t, F_2^S\right) = F_2^S(x,t)\exp\left[\int\frac{L_1(x)}{L_2(x)}dx\right].$$

## 2. (a) Complete and Particular Solutions

Since *U* and *V* are two independent solutions of equation (9) and if *α* and *β* are arbitrary constants, then $V = \alpha U + \beta$ may be taken as a complete solution [19-20] of equation (9). So, the complete solution

$$F_2^S(x,t)\exp\left[\int\frac{L_1(x)}{L_2(x)}dx\right] = \alpha t\exp\left[\frac{1}{A_f}\int\frac{L_1(x)}{L_2(x)}dx\right] + \beta \quad (10)$$

is a two-parameter family of surfaces. The one parameter family determined by taking $\beta = \alpha^2$ has equation

$$F_2^S(x,t)\exp\left[\int\frac{L_1(x)}{l_2(x)}dx\right] = \alpha t\exp\left[\frac{1}{A_f}\int\frac{1}{L_2(x)}dx\right] + \alpha^2. \quad (11)$$

Differentiating equation (11) with respect to *α*, we get $\alpha = -\frac{1}{2}t\exp\left[\frac{1}{A_f}\int\frac{1}{L_2(x)}dx\right]$. Putting the value of *α* in equation (11), we get

$$F_2^S(x,t) = -\frac{1}{4}t^2 \exp\left[\int\left(\frac{2}{A_f L_2(x)} - \frac{L_1(x)}{L_2(x)}\right)dx\right], \qquad (12)$$

which is merely a particular solution of the general solution. Now, defining

$$F_2^S(x,t_0) = -\frac{1}{4}t_0^2 \exp\left[\int\left(\frac{2}{A_f L_2(x)} - \frac{L_1(x)}{L_2(x)}\right)dx\right],$$ at $t = t_0$, where, $t_0 = \ln(Q_0^2/\Lambda^2)$ at any lower value $Q$ = $Q_0$, we get from equation (12)

$$F_2^S(x,t) = F_2^S(x,t_0)\left(\frac{t}{t_0}\right)^2, \qquad (13)$$

which gives the *t*-evolution of singlet structure function $F_2^S(x,t)$. Again defining,

$$F_2^S(x_0,t) = -\frac{1}{4}t^2 \exp\left[\int\left(\frac{2}{A_f L_2(x)} - \frac{L_1(x)}{L_2(x)}\right)dx\right]_{x=x_0},$$ we obtain from equation (12)

$$F_2^S(x,t) = F_2^S(x_0,t)\exp\left[\int_{x_0}^{x}\left(\frac{2}{A_f L_2(x)} - \frac{L_1(x)}{L_2(x)}\right)dx\right] \qquad (14)$$

which gives the *x*-evolution of singlet structure function $F_2^S(x,t)$. Proceeding in the same way, we get

$$F_2^{NS} = F_2^{NS}(x,t_0)\left(\frac{t}{t_0}\right)^2, \qquad (15)$$

which give the *t*-evolutions of non-singlet structure functions in LO. And also

$$F_2^{NS}(x,t) = F_2^{NS}(x_0,t)\exp\left[\int_{x_0}^{x}\left(\frac{2}{A_f Q(x)} - \frac{P(x)}{Q(x)}\right)dx\right], \qquad (16)$$

which give the *x*-evolutions of non-singlet structure functions in LO. Here,

$$P(x) = \frac{3}{2}A_1(x), \quad Q(x) = \frac{3}{2}A_3(x).$$

Proceeding exactly in the same way, from equations (3) and (4) we get

$$F_2^{S,NS}(x,t) = F_2^{S,NS}(x,t_0)\left(\frac{t^{(b/t+1)}}{t_0^{(b/t_0+1)}}\right)^2 \exp\left[2b\left(\frac{1}{t} - \frac{1}{t_0}\right)\right], \qquad (17)$$

$$F_2^S(x,t) = F_2^S(x_0,t)\exp\int_{x_0}^{x}\left[\frac{2}{a}\cdot\frac{1}{L_2(x)+T_0 M_2(x)} - \frac{L_1(x)+T_0 M_1(x)}{L_2(x)+T_0 M_2(x)}\right]dx, \qquad (18)$$

$$F_2^{NS}(x,t) = F_2^{NS}(x_0,t)\exp\int_{x_0}^{x}\left[\frac{2}{a}\cdot\frac{1}{A_3(x)+T_0 B_3(x)} - \frac{A_1(x)+T_0 B_1(x)}{A_3(x)+T_0 B_3(x)}\right]dx, \qquad (19)$$

which gives the $t$ and $x$-evolution of singlet and non-singlet structure functions in NLO where $a = 2/\beta_o$, $b = \beta_1/\beta_0^2$. We observe that in case of $t$-evolutions, if b tends to zero, then equation (17) tends to equation (13) and (14) respectively, i.e., solution of NLO equations goes to that of LO equations. Physically b tends to zero means number of flavours is high. Here, $M_1(x), M_2(x), B_1(x), B_2(x), B_3(x)$ and $B_4(x)$ are some functions of $x$ [5-7].

For all these particular solutions, taking $\beta = \alpha^2$. But if using $\beta = \alpha$ and differentiating with respect to $\alpha$ as before, the value of $\alpha$ can not be determined. In general, if taking $\beta = \alpha^y$, in the solutions the powers of $(t/t_0)$ and the numerators of the first term inside the integral sign be $y/(y-1)$ for $t$ and $x$-evolutions respectively in LO. Similarly the powers of $t^{b/t+1}/t_0^{b/t_0+1}$ and co-efficient of $b$ $(1/t-1/t_o)$ of exponential part in $t$-evolutions and the numerators of the first term inside the integral sign be $y/(y-1)$ for $x$-evolutions in NLO. Then if $y$ varies from minimum ($=2$) to maximum ($=\infty$) then $y/(y-1)$ varies from 2 to 1.

Deuteron, proton and neutron structure functions [21] can be written as

$$F_2^d(x, t) = (5/9) F_2^S(x, t), \tag{20}$$

$$F_2^p(x, t) = (5/18) F_2^S(x, t) + (3/18) F_2^{NS}(x, t), \tag{21}$$

$$F_2^n(x, t) = (5/18) F_2^S(x, t) - (3/18) F_2^{NS}(x, t). \tag{22}$$

Now using equations (13), (15) in equations (20-22) and (14) in equation (20) one obtains the $t$-evolutions of deuteron, proton, neutron and difference of proton and neutron and $x$-evolution of deuteron structure functions at low-$x$ as

$$F_2^{d,p,n}(x,t) = F_2^{d,p,n}(x,t_0)\left(\frac{t}{t_0}\right)^2, \tag{23}$$

$$F_2^p(x,t) - F_2^n(x,t) = [F_2^p(x,t_0) - F_2^n(x,t_0)](\frac{t}{t_0})^2, \tag{24}$$

$$F_2^d(x,t) = F_2^d(x_0,t)\exp\left[\int_{x_0}^{x}\left(\frac{2}{A_f L_2(x)} - \frac{L_1(x)}{L_2(x)}\right)dx\right] \tag{25}$$

in LO for $\beta = \alpha^2$. The corresponding results in NLO [6-7] for $\beta = \alpha^2$ are

$$F_2^{d,p,n}(x,t) = F_2^{d,p,n}(x,t_0)\left(\frac{t^{(b/t+1)}}{t_0^{(b/t_0+1)}}\right)^2 \exp\left[2b\left(\frac{1}{t} - \frac{1}{t_0}\right)\right], \tag{26}$$

$$F_2^p(x,t) - F_2^n(x,t) = [F_2^p(x,t_0) - F_2^p(x,t_0)]\left(\frac{t^{(b/t+1)}}{t_0^{(b/t_0+1)}}\right)^2 \exp\left[2b\left(\frac{1}{t} - \frac{1}{t_0}\right)\right], \tag{27}$$

$$F_2^d(x,t) = F_2^d(x_0,t)\exp\int_{x_0}^{x}\left[\frac{2}{a}\cdot\frac{1}{L_2(x)+T_0 M_2(x)} - \frac{L_1(x)+T_0 M_1(x)}{L_2(x)+T_0 M_2(x)}\right]dx, \qquad (28)$$

The determination of $x$-evolutions of proton and neutron structure functions like that of deuteron structure function is not suitable by this methodology; because to extract the $x$-evolution of proton and neutron structure functions, we are to put equations (14) and (16) in equations (21) and (22). But as the functions inside the integral sign of equations (14) and (16) are different, two separate the input functions $F_2^S(x_0, t)$ and $F_2^{NS}(x_0, t)$ are needed from the data points to extract the $x$-evolutions of the proton and neutron structure functions, which may contain large errors.

## 2. (b) Unique Solutions

Due to conservation of the electromagnetic current, $F_2$ must vanish as $Q^2$ goes to zero [21-22]. Since the value of $\Lambda$ is so small we can take at $Q = \Lambda$, $F_2^S(x, t) = 0$ due to conservation of the electromagnetic current [22]. This dynamical prediction agrees with most adhoc parameterizations and with the data [23]. Using this boundary condition in equation (10) we get $\beta = 0$ and

$$F_2^S(x,t) = \alpha t \exp\left[\int\left(\frac{1}{A_f L_2(x)} - \frac{L_1(x)}{L_2(x)}\right)dx\right]. \qquad (29)$$

Now, defining $F_2^S(x,t_0) = \alpha t_0 \exp\left[\int\left(\frac{1}{A_f L_2(x)} - \frac{L_1(x)}{L_2(x)}\right)dx\right]$, at $t = t_0$, where $t_0 = \ln(Q_0^2/\Lambda^2)$ at any lower value $Q = Q_0$, we get from equations (29)

$$F_2^S(x,t) = F_2^S(x,t_0)\left(\frac{t}{t_0}\right), \qquad (30)$$

which gives the $t$-evolutions of singlet structure function $F_2^S(x, t)$ in LO. Proceeding in the same way we get

$$F_2^S(x,t) = F_2^S(x_0,t)\exp\left[\int_{x_0}^{x}\left(\frac{1}{A_f L_2(x)} - \frac{L_1(x)}{L_2(x)}\right)dx\right] \qquad (31)$$

which gives the $x$-evolutions of singlet structure function $F_2^S(x, t)$ in LO. Similarly, for non-singlet structure functions

$$F_2^{NS}(x,t) = F_2^{NS}(x,t_0)\left(\frac{t}{t_0}\right), \qquad (32)$$

$$F_2^{NS}(x,t) = F_2^{NS}(x_0,t) \exp\left[\int_{x_0}^{x}\left(\frac{1}{A_f Q(x)} - \frac{P(x)}{Q(x)}\right)dx\right], \quad (33)$$

which give the *t* and *x*-evolutions of non-singlet structure functions in LO and

$$F_2^{S,NS}(x,t) = F_2^{S,NS}(x,t_0)\left(\frac{t^{(b/t+1)}}{t_0^{(b/t_0+1)}}\right)\exp\left[b\left(\frac{1}{t} - \frac{1}{t_0}\right)\right], \quad (34)$$

$$F_2^{S}(x,t) = F_2^{S}(x_0,t)\exp\int_{x_0}^{x}\left[\frac{1}{a}\cdot\frac{1}{L_2(x)+T_0 M_2(x)} - \frac{L_1(x)+T_0 M_1(x)}{L_2(x)+T_0 M_2(x)}\right]dx, \quad (35)$$

$$F_2^{NS}(x,t) = F_2^{NS}(x_0,t)\exp\int_{x_0}^{x}\left[\frac{1}{a}\cdot\frac{1}{A_5(x)+T_0 B_5(x)} - \frac{A_6(x)+T_0 B_6(x)}{A_5(x)+T_0 B_5(x)}\right]dx, \quad (36)$$

which give the *t* and *x*-evolutions of singlet and non-singlet structure functions in NLO.

Therefore corresponding results for *t*-evolution of deuteron, proton, neutron and difference of proton and neutron structure functions are

$$F_2^{d,p,n}(x,t) = F_2^{d,p,n}(x,t_0)\left(\frac{t}{t_0}\right), \quad (37)$$

$$F_2^{p}(x,t) - F_2^{n}(x,t) = [F_2^{p}(x,t_0) - F_2^{p}(x,t_0)]\left(\frac{t}{t_0}\right), \quad (38)$$

in LO and

$$F_2^{d,p,n}(x,t) = F_2^{d,p,n}(x,t_0)\left(\frac{t^{(b/t+1)}}{t_0^{(b/t_0+1)}}\right)\exp\left[b\left(\frac{1}{t} - \frac{1}{t_0}\right)\right], \quad (39)$$

$$F_2^{p}(x,t) - F_2^{n}(x,t) = [F_2^{p}(x,t_0) - F_2^{p}(x,t_0)]\left(\frac{t^{(b/t+1)}}{t_0^{(b/t_0+1)}}\right)\exp\left[b\left(\frac{1}{t} - \frac{1}{t_0}\right)\right], \quad (40)$$

in NLO. Again *x*-evolution of deuteron structure function in LO and NLO respectively are

$$F_2^{d}(x,t) = F_2^{d}(x_0,t)\exp\left[\int_{x_0}^{x}\left(\frac{1}{A_f L_2(x)} - \frac{L_1(x)}{L_2(x)}\right)dx\right], \quad (41)$$

$$F_2^{d}(x,t) = F_2^{d}(x_0,t)\exp\int_{x_0}^{x}\left[\frac{1}{a}\cdot\frac{1}{L_2(x)+T_0 M_2(x)} - \frac{L_1(x)+T_0 M_1(x)}{L_2(x)+T_0 M_2(x)}\right]dx. \quad (42)$$

Already we have mentioned that the determination of *x*-evolutions of proton and neutron structure functions like that of deuteron structure function is not suitable by this methodology. It is to be noted that unique solutions of evolution equations of different structure functions are same with particular solutions for maximum $y$ ($y = \infty$) in $\beta = \alpha^y$ relation.

## 2. (c) Approximate Solutions

It is to be noted that approximate solution of DGLAP evolution equation is obtained by considering $\alpha U + \beta V = 0$ instead of $V = \alpha U + \beta$ in equation (10) and the results [9-12] are same with the result of unique solutions.

## 2.(d) Regge behaviour

Using the Regge behaviour of singlet and non-singlet structure functions [12-13] as $F_2^S(x,t) = T_1(t)x^{-\lambda_S}$ and $F_2^{NS}(x,t) = T_2(t)x^{-\lambda_{NS}}$, singlet structure functions $F_2^S(x/\omega,t)$ and non-singlet structure function $F_2^S(x/\omega,t)$ can be approximated as

$$F_2^S(x/\omega,t) = T_1(t)\omega^{\lambda_S} x^{-\lambda_S} = F_2^S(x,t)\omega^{\lambda_S}, \tag{43}$$

$$F_2^{NS}(x/\omega,t) = F_2^{NS}(x,t)\omega^{\lambda_{NS}}, \tag{44}$$

where $T_1(t)$ and $T_2(t)$ are functions of $t$, and $\lambda_S$ and $\lambda_{NS}$ are the Regge intercepts for singlet and non-singlet structure functions respectively.

Using Regge behaviour of structure function and the relation between gluon and singlet structure function (equation (8)) in equation (1) one obtains the following form of equation

$$\frac{\partial F_2^S(x,t)}{\partial t} - \frac{F_2^S}{t} R_1(x) = 0, \tag{45}$$

where, $R_1(x)$ is a some function of $x$ [13].

Equation (45) can be solve as

$$F_2^S(x,t) = K t^{R_1(x)}, \tag{46}$$

where K is a integration constant.

From equation (46), the $t$ and $x$-evolutions of singlet structure function in LO can be obtained as

$$F_2^S(x,t) = F_2^S(x,t_0) \left(\frac{t}{t_0}\right)^{R_1(x)}, \tag{47}$$

$$F_2^S(x,t) = F_2^S(x_0,t) t^{\{R_1(x) - R_1(x_0)\}}. \tag{48}$$

Proceeding in the same way, $t$ and $x$-evolutions of non-singlet structure function in LO can be obtained as

$$F_2^{NS}(x,t) = F_2^{NS}(x_0,t)\left(\frac{t}{t_0}\right)^{R_2(x)} \tag{49}$$

$$F_2^{NS}(x,t) = F_2^{NS}(x_0,t)\, t^{\{R_2(x)-R_2(x_0)\}}. \tag{50}$$

where, $R_2(x)$ is a some function of $x$ [13].

The $t$ and $x$-evolution of singlet and non-singlet structure functions corresponding to NLO are respectively

$$F_2^S(x,t) = F_2^S(x,t_0)\left(\frac{t}{t_0}\right)^{R_3(x)}, \tag{51}$$

$$F_2^{NS}(x,t) = F_2^{NS}(x,t_0)\left(\frac{t}{t_0}\right)^{R_4(x)}, \tag{52}$$

$$F_2^S(x,t) = F_2^S(x_0,t)\, t^{\{R_3(x)-R_3(x_0)\}}, \tag{53}$$

$$F_2^{NS}(x,t) = F_2^{NS}(x_0,t)\, t^{\{R_4(x)-R_4(x_0)\}}, \tag{54}$$

where, $R_3(x)$ and $R_4(x)$ are some functions of $x$ [13].

Now using equations (47), (49) and (48), (50) in equations (20), (21) and (22), the $t$ and $x$-evolutions of deuteron, proton and neutron structure functions at low-$x$ can be obtained as

$$F_2^d(x,t) = F_2^d(x,t_0)\left(\frac{t}{t_0}\right)^{R_1(x)}, \tag{55}$$

$$F_2^d(x,t) = F_2^d(x_0,t)\, t^{\{R_1(x)-R_1(x_0)\}}, \tag{56}$$

$$F_2^P(x,t) = F_2^P(x,t_0)\frac{5t^{R_1(x)}+3t^{R_2(x)}}{5t_0^{R_1(x)}+3t_0^{R_2(x)}}, \tag{57}$$

$$F_2^P(x,t) = F_2^P(x_0,t)\frac{5t^{R_1(x)}+3t^{R_2(x)}}{5t^{R_1(x_0)}+3t^{R_2(x_0)}}, \tag{58}$$

$$F_2^n(x,t) = F_2^n(x,t_0)\frac{5t^{R_1(x)}-3t^{R_2(x)}}{5t_0^{R_1(x)}-3t_0^{R_2(x)}}, \tag{59}$$

and

$$F_2^n(x,t) = F_2^n(x_0,t) \frac{5t^{R_1(x)} - 3t^{R_2(x)}}{5t^{R_1(x_0)} - 3t^{R_2(x_0)}}. \tag{60}$$

in LO. The corresponding results in NLO are

$$F_2^d(x,t) = F_2^d(x,t_0) \left(\frac{t}{t_0}\right)^{R_3(x)}, \tag{61}$$

$$F_2^d(x,t) = F_2^d(x_0,t) \, t^{\{R_3(x) - R_3(x_0)\}}, \tag{62}$$

$$F_2^P(x,t) = F_2^P(x,t_0) \frac{5t^{R_3(x)} + 3t^{R_4(x)}}{5t_0^{R_3(x)} + 3t_0^{R_4(x)}}, \tag{63}$$

$$F_2^P(x,t) = F_2^P(x_0,t) \frac{5t^{R_3(x)} + 3t^{R_4(x)}}{5t^{R_3(x_0)} + 3t^{R_4(x_0)}}, \tag{64}$$

$$F_2^n(x,t) = F_2^n(x,t_0) \frac{5t^{R_3(x)} - 3t^{R_4(x)}}{5t_0^{R_3(x)} - 3t_0^{R_4(x)}}, \tag{65}$$

and

$$F_2^n(x,t) = F_2^n(x_0,t) \frac{5t^{R_3(x)} - 3t^{R_4(x)}}{5t^{R_3(x_0)} - 3t^{R_4(x_0)}}. \tag{66}$$

It is to be noted that Taylor series expansion method can not be used to solve DGLAP evolution equations in regge behaviour of structure functions. Since in regge behavior, region of discussion is at very low-x, so boundary condition $F_2(x, t)$ at $x = 1$ also can not be used.

## 2. (e) Characteristic method

For method of characteristics, two new variables $S$ and $\tau$ used instead of $x$ and $t$ [8] in equation (10), such that

$$\frac{dt}{dS} = -t, \tag{67}$$

$$\frac{dx}{dS} = A_f L_2(x) \tag{68}$$

Therefore equation (10) can be written as

$$\frac{dF_2^S(S,\tau)}{dS} + L_1(S,\tau) F_2^S(S,\tau) = 0. \tag{69}$$

Solution of equation (69) is

$$F_2^S(S,\tau) = F_2^S(\tau) \left(\frac{t}{t_0}\right)^{L_1(S,\tau)}, \tag{70}$$

where $L_1(S, \tau) = A_f \cdot L_1(x)$ and $F_2^S(S, \tau) = F_2^S(\tau); S = 0, t = t_0$.

After changing the variable ($S$ and $\tau$) to the original variable ($x$ and $t$), the $t$ and $x$-evolution of singlet structure function in LO [8] can be obtained as

$$F_2^S(x,t) = F_2^S(x,t_0)\left(\frac{t}{t_0}\right)^{A_f L_1(x)}.\tag{71}$$

$$F_2^S(x,t) = F_2^S(x_0,t)\exp\left[-\int_{x_0}^{x}\frac{L_1(x)}{L_2(x)}dx\right].\tag{72}$$

Proceeding in the same way, $t$ and $x$ evolutions of non-singlet structure function can be obtained as

$$F_2^{NS}(x,t) = F_2^{NS}(x,t_0)\left(\frac{t}{t_0}\right)^{A_1(x)},\tag{73}$$

$$F_2^{NS}(x,t) = F_2^{NS}(x_0,t)\exp\left[-\int_{x_0}^{x}\frac{A_1(x)}{A_3(x)}dx\right].\tag{74}$$

Now using equations (71) and (72) in equations (20), $t$ and $x$-evolution of deuteron structure functions in LO can be obtained as

$$F_2^d(x,t) = F_2^d(x,t_0)\left(\frac{t}{t_0}\right)^{A_f L_1(x)},\tag{75}$$

$$F_2^d(x,t) = F_2^d(x_0,t)\exp\left[-\int_{x_0}^{x}\frac{L_1(x)}{L_2(x)}dx\right],\tag{76}$$

The corresponding results in NLO are

$$F_2^d(x,t) = F_2^d(x,t_0)\left(\frac{t}{t_0}\right)^{\frac{3}{2}A_f\left[L_1(x)+T_0 M_1(x)\right]},\tag{77}$$

$$F_2^d(x,t) = F_2^d(x_0,t)\exp\left[-\int_{x_0}^{x}\frac{L_1(x)+T_0 M_1(x)}{L_2(x)+T_0 M_2(x)}dx\right],\tag{78}$$

Since the equation (71) and (73) as well as (72) and (74) are not in the same form, so two separate the input functions $F_2^S(x_0, t)$ and $F_2^{NS}(x_0, t)$ are needed from the data points to extract the $t$ and $x$-evolution of proton and neutron structure function. So using equations (21) and (22), determination of evolutions of proton and neutron structure functions is not possible. In all the methods, for possible solutions in NLO, an extra assumption [7, 11] $\left(\frac{\alpha_s(t)}{2\pi}\right)^2 = T_0\left(\frac{\alpha_s(t)}{2\pi}\right) = T_0 T_1$ is to be introduced, where $T_0$ is a numerical parameter and $T_1 = \left(\frac{\alpha_s(t)}{2\pi}\right)$. By a suitable choice of $T_0$ we can reduce the error to a minimum.

## 3. (a) Results and Discussion for Particular, unique and Approximate solutions

Results of particular solutions [5-7] of $t$-evolution of deuteron, proton, neutron and difference of proton and neutron structure functions compared with the NMC [25] and HERA [26] low-$x$ and low-$Q^2$ data and results of $x$- evolution of deuteron structure functions with NMC low-$x$ and low-$Q^2$ data. In case

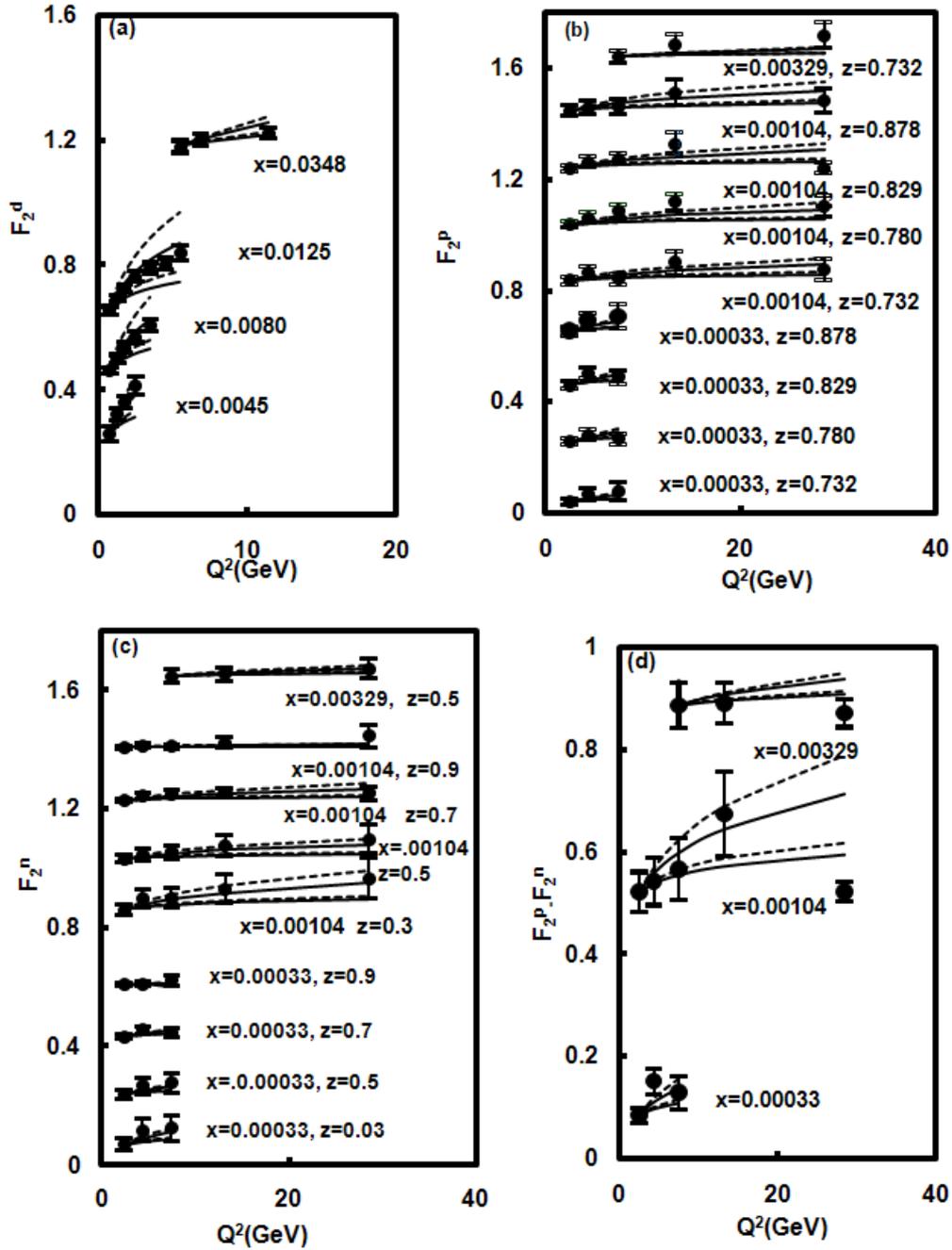

**Fig.1 (a-d):** Results of $t$-evolutions of deuteron, proton, neutron and difference of proton and neutron structure functions (dashed lines for LO and solid lines for NLO) for the representative values of $x$ in LO and NLO for NMC data. For convenience, value of each data point is increased by adding $0.2i$, where $i$ = 0, 1, 2, 3 ... are the numberings of curves counting from the bottom of the lowermost curve as the 0-th order. Data points at lowest-$Q^2$ values in the figures are taken as input.

of HERA data, proton and neutron structure functions are measured in the range $2 \leq Q^2 \leq 50$ GeV$^2$. Moreover, here $P_T \leq 200$ MeV, where $P_T$ is the transverse momentum of the final state baryon. In case of NMC data, proton and neutron structure functions are measured in the range $0.75 \leq Q^2 \leq 27$ GeV$^2$. We consider number of flavours $n_f = 4$.

In fig.1(a-d), represents results of t-evolutions of deuteron, proton, neutron and difference of proton and neutron structure functions (solid lines) for the representative values of $x$ given in the figures for $y = 2$ (upper solid lines) and $y$ maximum (lower solid lines) in $\beta = \alpha^y$ relation in NLO. Data points at lowest-$Q^2$ values in the figures are taken as input to test the evolution equation. Agreement with the data [25-26] is good. The same figures, represents the results of *t*-evolutions of deuteron, proton, neutron and difference of proton and neutron structure functions (dashed lines) for the particular solutions in LO. Here, upper dashed lines for $y = 2$ and lower dashed lines for $y$ maximum in $\beta = \alpha^y$ relation. We observe that *t*-evolutions are slightly steeper in LO calculations than those of NLO. But differences in results for proton and neutron structure functions are smaller and NLO results for $y = 2$ are of better agreement with experimental data in general.

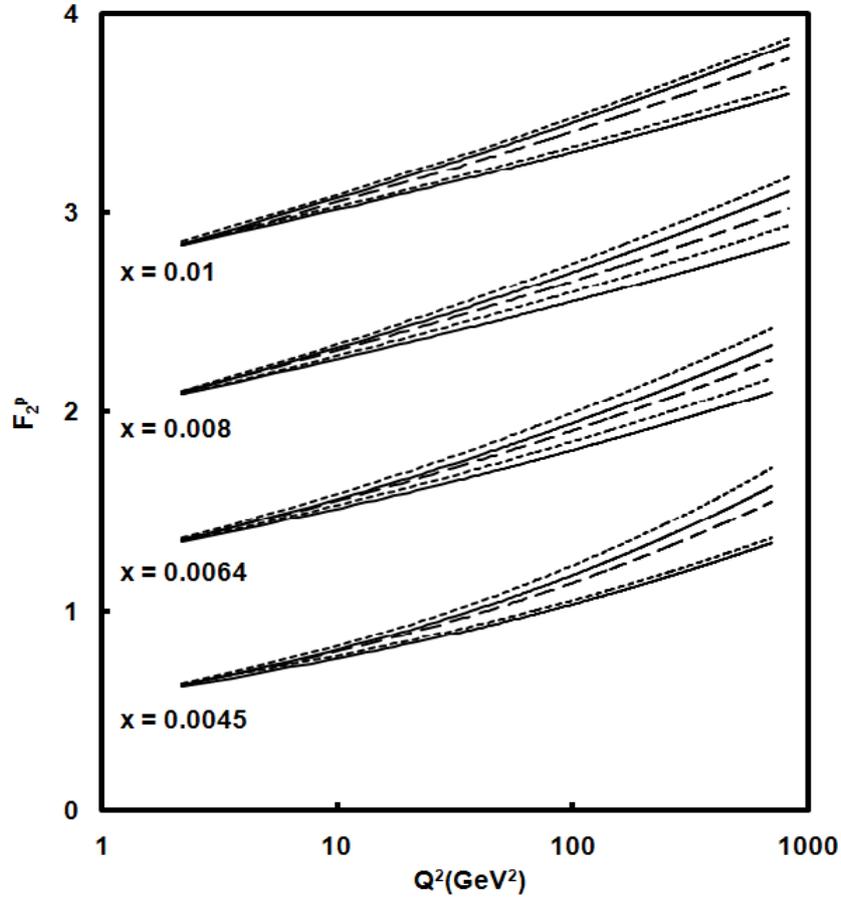

**Fig.2:** Results of *t*-evolutions of proton structure functions $F_2^p$ (dashed lines for LO and solid lines for NLO) with recent global paramatrization (long dashed lines) for the representative values of *x* given in the figures. Data points at lowest-$Q^2$ values in the figures are taken as input. For convenience, value of each data point is increased by adding $0.5i$, where $i = 0, 1, 2, 3, ...$ are the numberings of curves counting from the bottom of the lowermost curve as the 0-th order.

In fig.2, we compare our results of *t*-evolutions of proton structure functions $F_2^p$ (solid lines) with recent global parameterization [27] (long dashed lines) for the representative values of *x* given in the figures for *y* = 2 (upper solid lines) and *y* maximum (lower solid lines) in $\beta = \alpha^y$ relation in NLO. Data points at lowest-$Q^2$ values in the figures are taken as input to test the evolution equation. In the same figure, we also plot the results of *t*-evolutions of proton structure functions $F_2^p$ (dashed lines) for the particular solutions in LO. Here, upper dashed lines for *y* = 2 and lower dashed lines for *y* maximum in $\beta = \alpha^y$ relation. We observe that *t*-evolutions are slightly steeper in LO calculations than those of NLO. Agreement with the NLO results is found to be better than with the LO results.

Unique and approximate solutions of *t*-evolution for structure functions are same with particular solutions for *y* maximum ($y = \infty$) in $\beta = \alpha^y$ relation in LO and NLO.

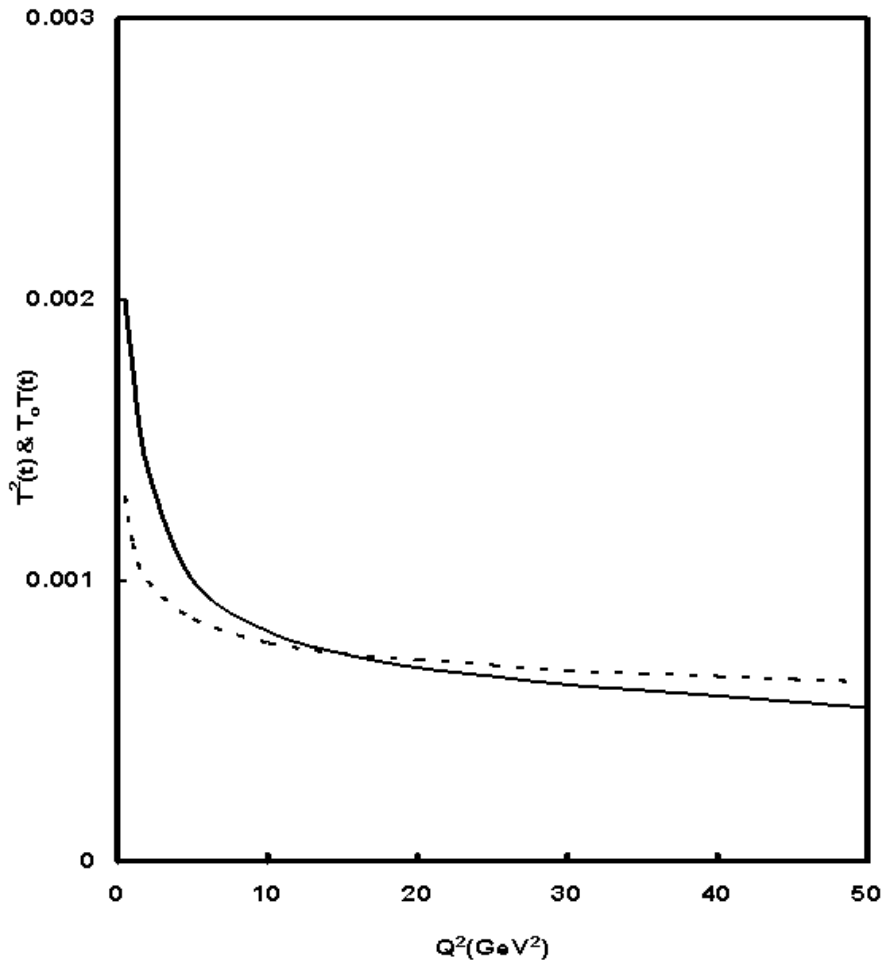

**Fig.3:** $T(t)^2$ and $T_0 T(t)$, where $T(t) = \alpha_s(t)/2\pi$ against $Q^2$ in the $Q^2$ range $0 \leq Q^2 \leq 50$ GeV$^2$.

In fig.3, we plot $T(t)^2$ and $T_0 T(t)$, where $T(t) = \alpha_s(t)/2\pi$ against $Q^2$ in the $Q^2$ range $0 \leq Q^2 \leq 50$ GeV$^2$ as required by our data used. Though the explicit value of $T_0$ is not necessary in calculating *t*- evolution of, yet we observe that for $T_0 = 0.108$, errors become minimum in the $Q^2$ range $0 \leq Q^2 \leq 50$ GeV$^2$.

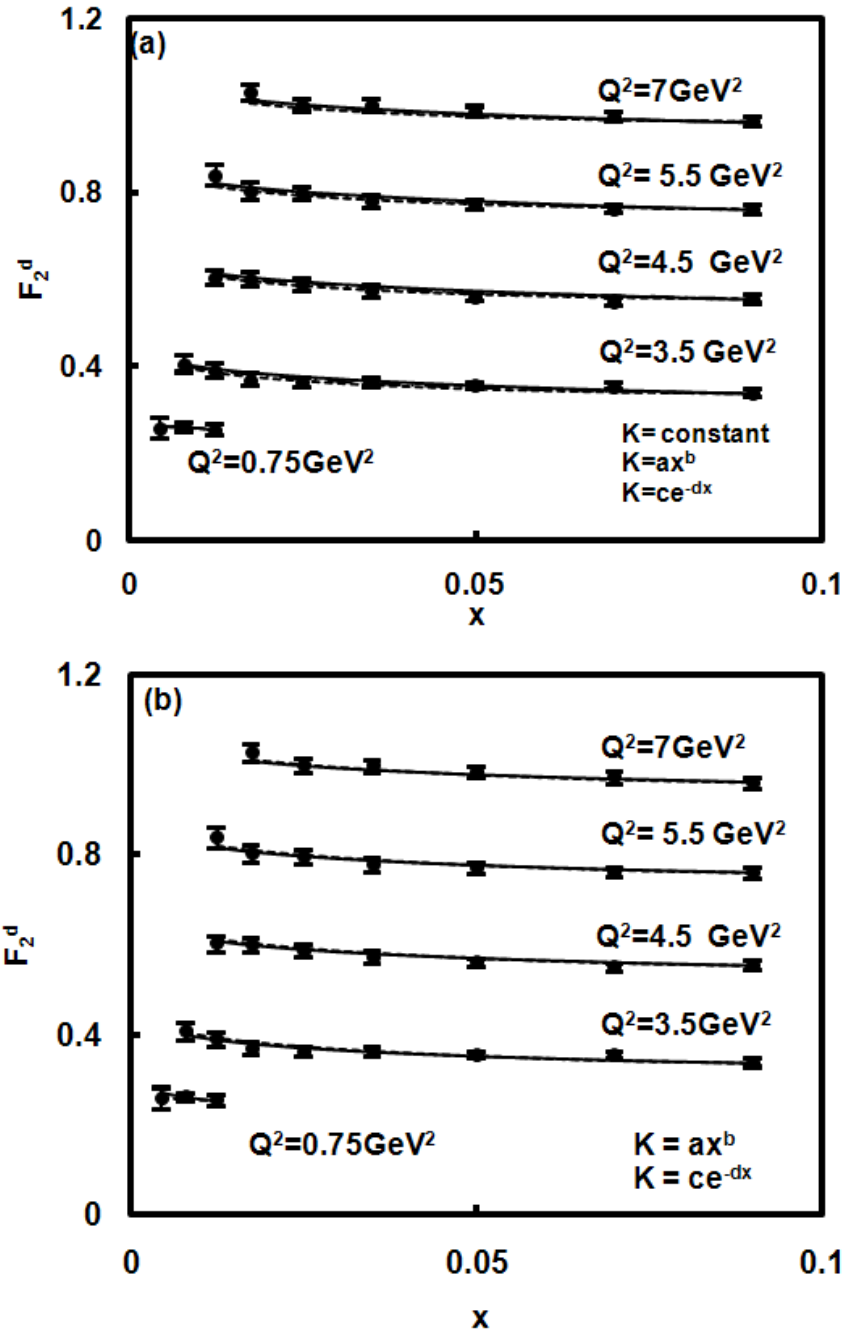

**Fig. 4(a-b):** Results of x-distribution of deuteron structure functions $F_2^d$ in LO for $K(x) = k$ (constant) (solid lines), $K(x) = ax^b$ (dashed lines) and for $K(x) = ce^{-dx}$ (dotted lines), where $k = 4.5$, $a = 4.5$, $b = 0.01$, $c = 5$, $b = 1$ and in NLO for $K(x) = ax^b$ (solid lines), and for $K(x) = ce^{-dx}$ (dotted lines), where $a = 5.5$, $b = 0.016$, $c = 0.28$, and $d = -3.8$ and for representative values of $Q^2$ given in each figure, and compare them with NMC deuteron low-x low-$Q^2$ data. In each the data point for x-value just below 0.1 has been taken as input $F_2^d(x_0, t)$. For convenience, value of each data point is increased by adding $0.2i$, where $i = 0, 1, 2, 3, \ldots$ are the numberings of curves counting from the bottom of the lowermost curve as the 0-th order.

In figs.4 (a-b), represents results of x-distribution of deuteron structure functions $F_2^d$ in LO (fig. 4(a)) for $K(x) = k$ (constant) (solid lines), $K(x) = ax^b$ (dashed lines) and for $K(x) = ce^{-dx}$ (dotted lines), and in

NLO (fig.4(b)) for $K(x) = ax^b$ (solid lines) and for $K(x) = ce^{-dx}$ (dashed lines) where $a$, $b$, $c$ and $d$ are constants and for representative values of $Q^2$ given in each figure , and compare them with NMC deuteron low-$x$ low-$Q^2$ data [25]. In each data point for $x$-value just below $0.1$ has been taken as input $F_2^d(x_0, t)$. In case of LO, agreement of the results with experimental data is good at $k = 4.5$, $a = 4.5$, $b = 0.01$, $c = 5$, $d = 1$. For $x$-evolutions of deuteron structure function, results of unique solutions and results of particular solutions have not any significance difference in LO [6]. In case of NLO, agreement of the result with experimental data is found to be excellent at $a = 10$, $b = 0.016$, $c = 0.5$, $d = -3.8$ for $y$ minimum ($y = 2$) and $a = 5.5$, $b = 0.016$, $c = 0.28$, $d = -3.8$ for $y$ maximum ($y = \infty$) in relation $\beta = \alpha^y$. But agreement of the results with experimental data is found to be very poor for any constant value of $k$. Therefore we do not present our result at $K(x) = k$ in NLO.

## 3. (b) Results and Discussion for Regge behavior

Though the results of regge behavior has been discussed elsewhere [13], here we mention some important point.

Nature of results of $t$ and $x$- evolution of structure functions is same with the results of particular solutions and agreement of the result with experimental data and global parameterization is good. In all the result from experimental as well as global fits, it is seen that structure functions increases when $x$ decreases and $Q^2$ increases for fixed values of $Q^2$ and $x$ respectively. But the results of $t$ and $x$- evolution of structure functions are not unique which depend on various parameters $K(x)$, $\lambda_S$, $\lambda_{NS}$ in LO and $K(x)$, $\lambda_S$, $\lambda_{NS}$ and $T_0$ in NLO.

$K(x)$ comes from the relation between gluon and singlet structure function, which is a function of $x$. Here also taking some simple standard functional forms of $K(x)$ which are same with the particular solutions i.e., $K(x) = k$, $ax^b$, and $ce^{dx}$. Explicit form of $K(x)$ can actually be obtained only by solving coupled DGLAP evolution equations for singlet and gluon structure functions considering regge behaviour. $\lambda_S$ and $\lambda_{NS}$ are regge intercepts for singlet and non-singlet structure functions and $T_0$ is a numerical parameter.

It is observed that result is sensitive to arbitrary parameters $k$, $a$, $b$, $c$, $d$ and $\lambda_S$, $\lambda_{NS}$, $T_0$ in $t$ and $x$-evolutions.

## 3.(c) Results and Discussion for Characteristic methods

Nature of results [8] of characteristic method for $t$ and $x$- evolution of structure functions is same with the results of particular solutions and agreement of the result with experimental data and global parameterization is good. In all the result from experimental as well as global fits, it is seen that structure functions increases when $x$ decreases and $Q^2$ increases for fixed values of $Q^2$ and $x$ respectively. But the results of $t$ and $x$- evolution of structure functions are not unique which depend on parameters $K(x)$ in LO and $K(x)$ and $T_0$ in NLO.

Here also taking some simple standard functional forms of $K(x)$ which are same with the particular solutions i.e., $K(x) = k$, $ax^b$, and $ce^{dx}$. $T_0$ is a numerical parameter.

It is observed that result is sensitive to arbitrary parameters $k$, $a$, $b$, $c$, $d$ and $T_0$ in $t$ and $x$-evolutions.

## Comparison of evolution results

The evolution results are discussed in section-3(a) for the particular, unique and approximate methods. Particular and unique solutions of singlet and non-singlet structure functions at low-$x$ are obtain using by Taylor's expansion method from GLDAP evolution equations and derive $t$-evolution for deuteron, proton, neutron and difference of proton and neutron structure functions and $x$-evolutions of deuteron structure functions and compare them with global data and parameterizations with satisfactory phenomenological success. Particular solutions of DGLAP evolution equations in LO and NLO obtained by that methodology were not unique and so the $t$- evolutions of deuteron, proton and neutron structure functions, and $x$-evolution of deuteron structure function obtained by this methodology were not unique. Thus by this methodology, instead of having a single solution we arrive a band of solutions, of course the range for these solutions is reasonably narrow.

In case of unique solutions, it has been observed that though we have derived a unique $t$-evolution for deuteron, proton, neutron, difference and ratio of proton and neutron structure functions in LO and NLO, yet we can not establish a completely unique $x$-evolution for deuteron structure function in LO and NLO due to the relation $K(x)$ between singlet and gluon structure functions and an adhoc parameter $T_0$ in NLO. This parameter does not effect in the results of $t$-evolution of structure functions. $K(x)$ may be in the forms of a constant, an exponential function or a power function and they can equally produce required $x$-distribution of deuteron structure functions. But unlike many parameter arbitrary input $x$-distribution functions generally used in the literature, these methods require only one or two such parameters. Unique solutions are obtain using by boundary condition, structure function $F_2 = 0$ at $x = 1$. Unique and approximate solutions of $t$ and $x$-evolution for structure functions are same with particular solutions for $y$ maximum ($y = \infty$) in $\beta = \alpha^y$ relation in LO and NLO. In all the result from experimental as well as global fits, it is seen that deuteron structure functions increases when $x$ decreases and $Q^2$ increases for fixed values of $Q^2$ and $x$ respectively, and proton, neutron, difference of proton and neutron structure functions increases when $Q^2$ increases for fixed value of $x$.

It is to be noted that the determination of *x*-evolutions of proton and neutron structure functions like that of deuteron structure function is not suitable by this methodology; because to extract the *x*-evolution of proton and neutron structure functions, two separate singlet input function $F_2^S(x_0, t)$ and non-singlet input functions $F_2^{NS}(x_0, t)$ are needed from the data points to extract the *x*-evolutions of the proton and neutron structure functions, which may contain large errors.

The evolution results are discussed in section-3(b) for the Regge behavior of structure functions. DGLAP evolution equations in LO and NLO have solved by considering Regge behavior of singlet and non-singlet structure functions at low-*x* and derive *t* and *x*-evolutions of various structure functions. It has been observed that *t* and *x*-evolutions for deuteron, proton and neutron structure functions in LO and NLO are not unique due to the relation $K(x)$ between singlet and gluon structure functions, Regge intercept $\lambda_S$, $\lambda_{NS}$ and an adhoc parameter $T_0$ in NLO. Where $\lambda_S$ and $\lambda_{NS}$ are the Regge intercepts for singlet and non-singlet structure functions respectively. $K(x)$ may be in the forms of a constant, an exponential function or a power function and they can equally produce required *t* and *x*-distribution of proton and deuteron structure functions. Explicit form of $K(x)$ can actually be obtained only by solving coupled DGLAP evolution equations for singlet and gluon structure functions considering regge behaviuor of structure functions [13]. On the other hand, we observed that the Taylor expansion method can not be used to solve DGLAP evolution equations considering regge behavior of structure functions. In all the result from experimental as well as global fits, it is seen that deuteron and proton structure functions increases when *x* decreases and $Q^2$ increases for fixed values of $Q^2$ and *x* respectively.

The evolution results are discussed in section-3(c) for the characteristic method. The solutions of singlet and non-singlet structure functions at low-*x* are obtained by using method of characteristic from GLDAP evolution equations and derive *t* and *x*-evolutions of deuteron structure functions. It has been observed that *t* and *x*-evolution for deuteron structure functions in LO and NLO are not unique due to the relation $K(x)$ between singlet and gluon structure functions and an adhoc parameter $T_0$ in NLO. $K(x)$ may be in the forms of a constant, an exponential function or a power function and they can equally produce required *t* and *x*-distribution of deuteron structure functions. In this method, boundary condition $F_2^S(S, \tau) = F_2^S(\tau)$; $t = t_0$, $x = \tau$ at $S = 0$ is used to obtain the solution. On the other hand, we observed that the Taylor expansion method can be used to solve DGLAP evolution equations in this method. In all the result from experimental as well as global fits, it is seen that deuteron and proton structure functions increases when *x* decreases and $Q^2$ increases for fixed values of $Q^2$ and *x* respectively.

It is to be noted that the determination of *t* and *x*-evolutions of proton and neutron structure functions like that of deuteron structure function is not suitable by this methodology; because to extract the *t* and *x*-evolution of proton and neutron structure functions, two separate singlet input function $F_2^S(x_0, t)$ and non-singlet input functions $F_2^{NS}(x_0, t)$ are needed from the data points to extract the *x*-evolutions of the proton and neutron structure functions, which may contain large errors.

Comparisons of these methods are summarized in Table-1.

Table-1

Summary of Comparisons of these evolution methods.

| Method | Advantage and Disadvantage |
|---|---|
| Particular | 1. Taylor expansion method can be used to solve DGLAP evolution equations.<br>2. Particular solutions of DGLAP evolution equations are not unique. We arrive at a band of solutions, of course the range for these solutions is reasonably narrow.<br>3. For *x*-evolutions of deuteron structure function, results for $y = 2$ and *y* maximum ($y = \infty$) in $\beta = \alpha^y$ relation do not have any significant difference.<br>4. The determination of *x*-evolutions of proton and neutron structure functions is not suitable by this methodology.<br>5. For possible solutions of DGLAP evolution equations in NLO, we introduce an adhoc numerical parameter $T_0$, which does not effect the results of t-evolution of structure functions.<br>6. Explicit form of $K(x)$ can not be obtained by solving coupled DGLAP evolution equations for singlet and gluon structure functions by this methodology.<br>7. In this method, boundary condition is not used to solve the DGLAP evolution equations. |
| Unique | 1. Taylor expansion method can be used to solve DGLAP evolution equations.<br>2. *t*-evolution of structure functions in LO and NLO are unique, but *x*- evolution for deuteron structure functions in LO and NLO are not unique due to the relation $K(x)$ between singlet and gluon structure functions and an adhoc parameter $T_0$ in NLO.<br>3. Unique solutions of DGLAP evolution equations are same with the particular solutions for maximum *y* ($y = \infty$) in $\beta = \alpha^y$ relation.<br>4. The determination of *x*-evolutions of proton and neutron structure functions is not suitable by this methodology. |
|   | 5. For possible solutions of DGLAP evolution equations in NLO, we introduce an adhoc numerical parameter $T_0$, which does not effect the results of $t$-evolution of structure functions. |
|   | 6. Explicit form of $K(x)$ can not be obtained by solving coupled DGLAP evolution equations for singlet and gluon structure functions by this methodology. |
|   | 7. Boundary condition [$F_2$= 0 at $x$ =1] is used to solve the DGLAP evolution equations. |
| Approximate | 1. Same with unique solutions up to no.6. |
|   | 2. Boundary condition is not used to solve the DGLAP evolution equations. |
| Regge | 1. Taylor expansion method can not be used to solve DGLAP evolution equations. |
|   | 2. $t$ and $x$-evolution of different structure functions in LO and NLO are not unique due to the relation $K(x)$ between singlet and gluon structure functions, Regge intercepts $\lambda_S$, $\lambda_{NS}$ and an adhoc parameter $T_0$ in NLO. |
|   | 3. The determination of $x$-evolutions of deuteron, proton and neutron structure functions is suitable by this methodology. |
|   | 4. For possible solutions of DGLAP evolution equations in NLO, we introduce an adhoc numerical parameter $T_0$, which effects the results of $t$ and $x$-evolution of structure functions. |
|   | 5. Explicit form of $K(x)$ can be obtained by solving coupled DGLAP evolution equations for singlet and gluon structure functions considering regge behaviour of structure functions. |
|   | 6. Boundary condition is not used to solve the DGLAP evolution equations. |
| Characteristic method | 1. Taylor expansion method can be used to solve DGLAP evolution equations. |
|   | 2. $t$ and $x$-evolution for deuteron structure functions in LO and NLO are not unique due to the relation $K(x)$ between singlet and gluon structure functions and an adhoc parameter $T_0$ in NLO. |
|   | 3. The determinations of $t$ and $x$-evolutions of proton and neutron structure functions are not suitable by this methodology. |
|   | 4. For possible solutions of DGLAP evolution equations in NLO, we introduce an adhoc numerical parameter $T_0$, which effects the results of $t$ and $x$-evolution of structure functions. |
|   | 5. Explicit form of $K(x)$ can not be obtained by solving coupled DGLAP evolution equations for singlet and gluon structure functions by this methodology. |
|   | 7. Boundary condition [$F_2^S(S, \tau) = F_2^S(\tau)$; $t = t_0$, $x = \tau$ at S = 0.] is used to solve the DGLAP evolution equations. |